\documentstyle[prl,aps,epsfig]{revtex}

\begin{document}
\wideabs{ \title{Quantum key distribution  with  bright entangled
beams}

\author{Ch.~Silberhorn,  N.~Korolkova, and G.~Leuchs}

\address{Zentrum f\"ur Moderne Optik an  der
Universit\"at Erlangen-N\"urnberg, \\ Staudtstra{\ss}e 7/B2,
D-91058 Erlangen, Germany.}

\date{\today}
\maketitle

\begin{abstract}
We suggest a quantum cryptographic scheme using continuous
EPR-like correlations of bright optical beams. For binary key
encoding, the continuous information is discretized in a novel way
by associating a respective measurement, amplitude or phase, with
a bit value "1" or "0". The secure key distribution is guaranteed
by the quantum correlations. No pre-determined information is sent
through the quantum channel contributing to the security of the
system.
\end{abstract}

}

\narrowtext


Quantum key distribution (QKD) is the most advanced technology in
the field of quantum information processing. The conventional
arrangements use dichotomic quantum systems to realize the secure
information transfer (for a review see \cite{gisin01}). These
discrete systems have the advantage to be in principle loss
insensitive in terms of security. However, the generation process
for entangled photon pairs needed for QKD based on EPR
correlations is spontaneous and therefore probabilistic. This
limits the achievable data transmission rates.

A new development employs continuous variable systems
\cite{reid00,hillery00,ralph00a,cerf00}, such as intense light
fields, to obtain shorter key distribution times. The security
issues of continuous variable quantum cryptography have been
addressed \cite{ralph00b,gottesman00} and it was proven that the
secure key distribution can be achieved using continuous EPR-type
correlation or quantum squeezed states.

In this Letter we propose a new key distribution scheme based on
the quantum EPR-like correlations of conjugate continuous
variables. The main novel feature of the protocol
\cite{silberhorn00} is an assignment of a bit value to the type of
measurement. The binary bits are encoded by the choice to detect
either of two conjugate variables accomplished independently and
randomly by both communicating parties.  This serves as a
discretization of continuous information in the measurement
process. The coincidences in the choices of both parties are
revealed by testing the EPR-like correlations between the beams
and contribute to the key. Thus, in contrast to other continuous
variable systems \cite{reid00,hillery00,ralph00a,cerf00}, the
basis value is not predetermined but develops in measurements at
receiver and sender stations,
resembling the EPR-based Ekert protocol 
for discrete cryptographic systems. The detection of light
statistics performed by both communicating parties plays a
decisive role in the proposed scheme. It comprises bit encoding,
key sifting, monitoring of the disturbance in the quantum channel
and active control on timing and information flow during the
transmission \cite{silberhorn}.

The basic ingredient of the scheme are quantum correlations
between the amplitude $\hat X_j = \hat a^\dagger_j + \hat a_j $
and phase $Y_j = i (\hat a^\dagger_j - \hat a_j)$ quadratures of
bright beams $j=1,2$. Because of the high intensity of the optical
fields involved, we use the linearization approach throughout the
paper: $\hat X_{\rm j} = \langle X_{\rm j} \rangle + \delta \hat
X_{\rm j}$, $\hat Y_{\rm j} = \langle Y_{\rm j} \rangle + \delta
\hat Y_{\rm j}$. The entangled observables are then the quantum
uncertainties in the respective field quadratures. We start with
the definition of the relevant measured quantities and of the
conditions  for appling the two-mode correlations  as a quantum
resource. It can be done on the basis of the non-separability
criterion for continuous variables \cite{simon,duan}.

The Peres-Horodecki  criterion for continuous variables  provides
a sufficient condition for a Gaussian state to be non-separable
\cite{simon,duan}. It can be written in terms of sum or difference
squeezing variances \cite{reid} of amplitude and phase of two
beams:
\begin{eqnarray}
   V_{\rm sq}^\pm (X) = \frac{V(\delta \hat{X}_{1}
 \pm g\  \delta \hat{X}_{2})}{V( \hat{X}_{1, {\rm SN}} +
   g  \hat{X}_{2, {\rm SN}})}, \label{sq-amp} \\
   V_{\rm sq}^\mp(Y) = \frac{V(\delta \hat{Y}_{1}
   \mp g \ \delta \hat{Y}_{2})}{V( \hat{Y}_{1, {\rm SN}}+
  g \hat{Y}_{2, {\rm SN}})}
   \label{sq-ph}
\end{eqnarray}
where $ V (A)$ is the variance $\langle \hat A ^2 \rangle -
\langle \hat A \rangle ^2$ of an observable $\hat A$. The field
modes are denoted by the respective subscripts, ${\rm SN}$ labels
the shot noise limit for a corresponding beam, and $g$ is a
variable gain to minimize the variance \cite{reid}. In the
particular case of entirely symmetrical entangled beams the
optimal gain is calculated to be $g=1$ \cite{duan}. The
non-separability of the two-mode quantum state requires then
$V_{\rm sq}^+ (X) + V_{\rm sq}^- (Y) < 2$ \cite{simon,duan} and
the criterion is necessary and sufficient \cite{duan}. From here
on we use Eqn. (\ref{sq-amp}, \ref{sq-ph}) with the signs
corresponding to amplitude anti-correlations and phase
correlations. The two-mode non-separable state is said to be {\it
squeezed-state entangled} if  the following condition is satisfied
for the variances of conjugate variables in Eqs. (\ref{sq-amp},
\ref{sq-ph}) \cite{leuchs01}:
\begin{eqnarray}
V_{\rm sq}^+ (X)  < 1, \qquad V_{\rm sq}^- (Y)  < 1.
\label{sq-def}
\end{eqnarray}
Note,  that in contrast to the non-separability criterion, the
introduced  squeezed-state entanglement requires both variances of
conjugate variables to drop below the respective limit. This
requirement is crucial for the suggested cryptographic system and
ensures both the possibility to build up a binary key string and
the security of a transmission.

\begin{figure}[b]
\begin{center}
 \epsfxsize=0.85\columnwidth
 \epsfbox{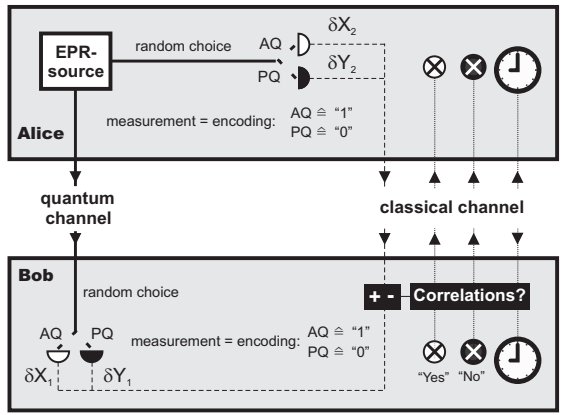}
\end{center} \caption{QKD with bright EPR-entangled
beams (see text).} \label{scheme}
\end{figure}


A key point for the QKD protocol is sum (difference)
measurement(\ref{sq-amp}, \ref{sq-ph}) testing for the correlation
in the amplitude and phase quadratures. It is used to determine a
bit value and it ensures an undisturbed transmission. Suppose
Alice and Bob both record the amplitude quadratures of their
respective EPR-beam.  In this case the detected quantum
uncertainties $\hat \delta X_1$ and $\hat \delta X_2$ are
anti-correlated \cite{silberhorn01}.  Bob tests  for
anti-correlations by recording the variance of the sum
(\ref{sq-amp}) of photo currents of his and Alice's measurement.
Note, however, that the time interval used to experimentally
determine the photo current statistics plays a crucial role for
the security of the protocol because it may allow for an
undetectable eavesdropping (see below). At this stage we explain
the protocol in terms of squeezing variances for the sake of
clarity in the presentation of main ideas. If there is a non-local
anti-correlation between $\delta X_1$ and $\delta X_2$, the sum
photo current will drop below the quantum limit, $ V_{\rm sq}^+
(X) < 1$, to the extent dependent on the quality of the EPR
source. Analogously, if there is non-local correlation between
$\delta Y_1$ and $\delta Y_2$ the difference photo current will
drop below the quantum limit $ V_{\rm sq}^- (Y) < 1$
(\ref{sq-ph}). The quality of the source is limited by the finite
degree of continuous quantum correlations $ V_{\rm sq} (X), V_{\rm
sq} (Y)$ (\ref{sq-amp}, \ref{sq-ph}), perfect correlation
requiring infinite energy resources. For the efficiency of
transmission, noise and losses in the quantum channel
 play a significant role. The net quality of both
the source and the channel has impact on the distance, on which
the quantum correlations are still reliably observable, on the
sensitivity to the disturbance by an eavesdropper and on possible
achievable bit rates.

 The obtained
constraint  $V_{\rm sq}(X)\ll 1$ and $V_{\rm sq}(Y) \ll 1$  serve
Bob  as a criterion for the generation of the sifted key and as a
test for eavesdropping. A measured normalized noise power of
$V_{\rm sq}(X)\ll 1$ at Bob's station delivers a bit value "$1$"
and $V_{\rm sq}(Y) \ll 1$ a bit value "$0$". The observation of
$V_{sq}(X), V_{sq}(Y)> 1$ means that both parties have  measured
different quadratures. These events are discarded. However, Alice
and Bob should keep controlling that the overall rate of the event
"no correlation" is statistically close to 50\% as is inherent to
the protocol ($X$ or $Y$ quadrature). $V_{sq}(X), V_{sq}(Y) < 1$
to an extent less then expected or no correlations in more than
50\% measurements reveals an unexpected disturbance in the line.
Note, that there is no need to communicate the obtained
constraints $V_{\rm sq}\ll 1$ to Alice.

The quantum key distribution protocol for squeezed-state entangled
bright beams based on the measurement of the EPR-like correlations
works as follows. The EPR-source is at Alice's station (Fig.
\ref{scheme}). Alice generates and distributes the entangled beams
keeping beam $1$ at her station and sending beam $2$ to Bob. The
relevant measured quantities are  the quadrature quantum
uncertainties which {\it a priori}  carry no information. To
establish the right timing of their recordings Alice and Bob have
to synchronize their clocks and agree upon a set of time intervals
$\Delta t_k$ in which they subdivide their measurements. Alice and
Bob proceed with a series of measurements.

The expected quality of quantum correlations  $V_{sq}(X),
V_{sq}(Y) < 1$  is determined experimentally as described above.
Alice and Bob start a key transmission by performing randomly and
independently measurements of either amplitude quadrature AQ or
phase quadrature PQ each. Hereby they keep recording: 1) their
photo currents ($\delta \hat X_j$,~$\delta \hat Y_j$), 2) the
respective time slots ($t_k$), and 3) the type of measurement
performed (AQ or PQ).

Bob and Alice use two classical communication channels to evaluate
the transmission results. Alice permanently keeps sending the
results of her measurements, the photo current containing $\delta
X_2^k$ or $\delta Y_2^k$ in the time slots $t_k$, to Bob via a
classical channel I. Bob performs the selection of "good" bits and
the security test. To generate the sifted key, he checks
correlations between the results of his measurements and the
results received from Alice by recording the variance
$V_{sq}^+(X)$ (\ref{sq-amp}) of the sum of the relevant photo
currents for his choice of AQ or the difference variance
$V_{sq}^-(Y)$ (\ref{sq-ph}) for PQ.

After evaluating his correlation measurements (Fig.~\ref{scheme}),
Bob publicly communicates to Alice  via the classical channel II
the time points $t_1, t_3,  ..., t_k$ when he detected
correlations (\ref{sq-def}).  The choice of the AQ~/~PQ
measurement is not disclosed. At this stage Alice  and Bob can
generate the common secret key. They pick up the  measurement
type (AQ/PQ) from their recordings at the time points $t_1, t_3,
..., t_k$ and build the secret key string by association: AQ =
bit value "1" and PQ = bit value "0". They discard the rest of
the data. The presence of Eve will be revealed by distortion of
the correlations or by events "no correlations" occurring
statistically  more frequent than $50\%$. This protocol is
summarized in Table 1.


The security of transmission against eavesdropping is guaranteed
by the sensitivity of the existing correlations to losses and by
the impossibility to measure both conjugate variables
simultaneously. The complete security analysis for the case of
continuous variables is non-trivial and lies beyond the scope of
the present Letter.  It will be considered elsewhere in terms of
mutual information and disturbance of transmission
\cite{luetkenhaus} and using noise characteristics like
signal-to-noise ratio \cite{ralph01,silberhorn}. Here we restrict
ourselves to the optical tap attack of an eavesdropper Eve and for
an idealized case of lossless quantum channel to illustrate the
main security mechanisms. The tapping beam splitter has a
transmissivity $\eta$.
\begin{figure}[b]
\begin{center}
 \epsfxsize=0.85\columnwidth
 \epsfbox{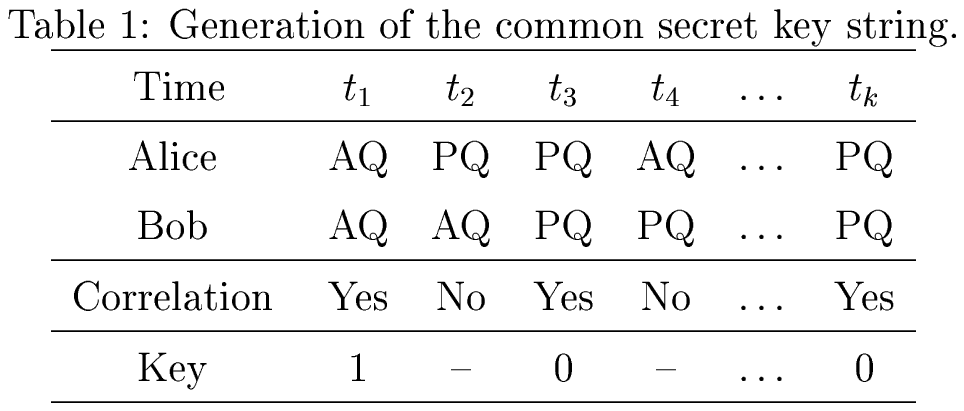}
\end{center}
\end{figure}

An eavesdropper Eve will attempt to figure out which quadrature
was measured by Alice by tapping the quantum channel and by trying
to relate these measurements to the photo currents travelling from
Alice to Bob through the classical channel. If Eve has decided to
detect the amplitude quadrature by tapping, she has at her
disposal the minus and plus channels:
\begin{eqnarray}
     V_{\rm sq}^\mp(X_{\rm E}; Z_{\rm A}) = \frac{V(\delta \hat{X}_{\rm E}
   \mp  g_{\rm E}\ \delta \hat{Z}_{\rm A})}{V( \hat{X}_{\rm E, {\rm SN}}+
   \hat{Z}_{\rm A, {\rm SN}})}, \  \  \hat Z_{\rm A}= \hat X_{\rm A},
\hat Y_{\rm A}. \label{EveDefSum}
\end{eqnarray}
Subscripts $\rm E, A$ denote the  quantum uncertainties, measured
respectively by Eve and by Alice, upper (lower) sign
 refers to the minus (plus) channel,  and
$g_{\rm E}$ is a variable gain used by Eve. To construct the
secret key, Eve must be able to distinguish between two possible
events: $\delta \hat Z_{\rm A}= \delta \hat X_{\rm A}$ or $\delta
\hat Z_{\rm A}= \delta \hat Y_{\rm A}$. An effective strategy for
Eve is to check the difference between her plus and minus channels
$\Delta= V_{\rm sq}^-(X_{\rm E};
 Z_{\rm A})-V_{\rm sq}^+(X_{\rm E};  Z_{\rm A})$. No difference between
recordings in these two channels reveals to Eve that she and Alice
have measured different quadratures. If Eve records a significant
difference $\Delta$, she knows that she and Alice have measured
the same quadrature:
 \begin{eqnarray}
 \displaystyle
  \Delta= V_{\rm sq}^-(X_E;
X_{\rm A})-V_{\rm sq}^+(X_E;  X_{\rm A}) = \label{EveDelta} \\
 g_{\rm E}\ \sqrt{ (1-\eta)}\  \left [
V_{\rm sq}^+(X) + V_{\rm sq}^-(X) \right ]. \nonumber
\end{eqnarray}
Here $V_{\rm sq}^{+} (X)$  is the normalized sum photo current
noise for the amplitude quadratures measured by Alice and Bob
during the undisturbed transmission. It is given by the squeezing
variance [Eq.~(\ref{sq-amp})] with the optimal gain $g=g_{\rm
sq}=1$. $V_{\rm sq}^{-} (X)$ [Eq.~(\ref{sq-amp})] is the
difference photo current noise which is recorded in Bob's minus
channel for an amplitude measurement. If the beams of Alice and
Bob are anti-correlated in the amplitude quadrature, the variance
$V_{\rm sq}^{+} (X)$ is well below unity. Due to the quantum
penalty the complimentary variance  $V_{\rm sq}^{-} (X)$ exhibits
then substantial excess noise. Equation (\ref{EveDelta}) thus
shows that though Eve can split off a small fraction of the signal
and process arbitrarily her measurement results which are
classical photo currents, she will be limited by  inherent noise
present in the signal. Eve can amplify her signal using the
electronic gain $g_{\rm E}$ but it will not improve the
signal-to-noise ratio of the detected light field.

We discuss now which means Alice and Bob have at their disposal to
reveal the malicious disturbance caused by Eve in the quantum
channel. First, we review another criterion for quantum EPR-like
correlations introduced by Reid and Drummond \cite{reid}. For the
discussions about this EPR condition and about the nonseparability
criterion see \cite{criteria,leuchs01} and references therein. The
EPR criterion refers to the demonstration of the EPR paradox for
continuous variables and specifies the ability to infer "at a
distance" either of the two non-commuting signal observables with
a precision below the vacuum noise level of the signal beam
\cite{reid}. The relevant inference errors \cite{reid} at the
optimal gain are the conditional variances:
\begin{eqnarray}
   V_{\rm cond}^\pm(X_1 \vert X_2)=  \frac{V(\delta \hat{X}_{1}
  \pm  g\ \delta \hat{X}_{2})}{V( \hat{X}_{1, {\rm SN}} )} \label{epr-amp} \\
      V_{\rm cond}^\mp (Y_1\vert Y_2)=  \frac{V(\delta \hat{Y}_{1}
  \mp g\ \delta \hat{Y}_{2})}{V( \hat{Y}_{1,{\rm SN}} )}
   \label{epr-ph}.
\end{eqnarray}
The demonstration of the EPR paradox for continuous variables
\cite{reid,ou,silberhorn01} corresponds to:
\begin{eqnarray}
V_{\rm cond}^+ (X_1\vert X_2) \ V_{\rm cond}^-(Y_1\vert Y_2) < 1.
\label{epr-merit}
\end{eqnarray}
An interesting tool to control the quantum channel is the variable
gain $g$ in definition of these conditional variances  $V_{\rm
cond}^\pm(X_1 \vert X_2)$ [Eqn.~(\ref{sq-amp}, \ref{epr-amp})].

Let us consider first the undisturbed transmission with an example
of the amplitude measurement performed by Bob.  Even for
symmetrical squeezed-state entangled beams the optimal gain to
minimize the conditional variance $V_{\rm cond}^+(X_1 \vert X_2)$
[Eq.~(\ref{epr-amp})] differs from unity. It can be expressed as:
\begin{eqnarray}
\displaystyle
   g_{\rm cond} = \frac{V_{\rm sq}^{-} (X) - V_{\rm sq}^{+} (X)}
   {V_{\rm sq}^{-} (X) + V_{\rm sq}^{+} (X)}\ .
   \label{gain}
    \end{eqnarray}
 With $V_{\rm sq}^{+} (X) \rightarrow 0$,
the noise variance in the minus channel $V_{\rm sq}^{-} (X)
\rightarrow \infty$ and the optimal gain for $V_{\rm cond}^+(X_1
\vert X_2)$ is also approaching unity $g_{\rm cond} \rightarrow
1$, like the optimal gain for the squeezing variances
[Eqn.~(\ref{sq-amp}, \ref{epr-amp})].

Consider now, how the invasion of an eavesdropper is reflected in
the measurements at Bob's station. The sum photo current measured
by Bob, $ V_{\rm sq}^{+} (X; \eta)$, becomes more noisy in the
presence of Eve:
\begin{eqnarray}
\displaystyle
   V_{\rm sq}^{+} (X; \eta)& = & \frac{(1+\sqrt{\eta})^2}{4}\ V_{\rm sq}^{+} (X)
     \label{AliceEve} \\
 & + & \frac{(1-\sqrt{\eta} )^2}{4}\ V_{\rm sq}^{-} (X)
    + \frac{1-\eta}{2} \nonumber
    \end{eqnarray}
 containing $V_{\rm sq}^{-} (X) \gg 1 $. Analogously, the signal in Bob's
minus channel, i.~e. the variance $V_{\rm sq}^{-} (X; \eta)$, will
be also changed, both plus and minus channel approaching the same
limit. Note that Eve should be cautious enough to keep the
classical amplitude of Bob's signal unchanged. Bob uses,
therefore, the unchanged value of the shot noise level to which
the measured noise variances are normalized to obtain $V_{\rm sq}$
corresponding to $g_{\rm sq}=1$.

The modified noise variances in the plus and minus channel $V_{\rm
sq}^{\pm} (X; \eta)$ will be reflected in   the optimal gain to
minimize the conditional variance $g_{\rm cond}$ (\ref{gain}):
\begin{eqnarray}
\displaystyle
   g_{\rm cond} (\eta)= \frac{V_{\rm sq}^{-} (X) -
    V_{\rm sq}^{+} (X)}
   {V_{\rm sq}^{-} (X) + V_{\rm sq}^{+} (X)}\ \sqrt{\eta} \ .
   \label{gainEta}
    \end{eqnarray}
 This gain also minimizes the observed unnormalized noise
variance $V(\delta \hat{X}_{1}
 \pm g\  \delta \hat{X}_{2})$.
 If Bob monitors $g_{\rm cond} (\eta)$ (\ref{gainEta})
in his measurements,
 he can easily infer $\eta \neq 1$ in the
quantum channel.

An important issue is the finite time for the confident
experimental determination of the squeezing variance. To gain some
partial information, when decoding $V_{\rm sq}$ Eve might go for
less time  than both legitimate communicating parties, accepting
less confidence in determining the variance. She will tap the
signal for a fraction of Bob's measurement time and hence will
introduce less disturbance as expected for a given beam splitting
ratio. The losses in the channel  reduce the correlations and
enhance the time needed for the determination of the variance with
sufficient precision. If Eve is tapping close to Alice, where the
impact of losses is still negligible, she can additionally profit
from less time needed for her measurement compared to that of Bob
with a given confidence level. The optimum strategy for Alice and
Bob seems to be to operate with as short a measurement time as
possible, ultimately with single measurements. The above
statistical analysis in terms of variances should therefore be
extended to cope with single shot measurements. This, however, is
beyond the scope of discussion presented here and will be
considered in detail elsewhere.

To summarize, the  scheme presented here possesses several  novel
features and shows the strong sides of continuous variable
cryptography. The effect of losses on the maximum possible
transmission distance will have to be studied further. The bit
value is encoded by the type of measurement, i. e. by the choice
of measured observable amplitude or phase. The information on the
key is thus emerging only "a posteriori", at sender and receiver
stations. One of the advantages of the presented scheme is the
high value of the achievable effective bit rates. For example, for
the pulsed EPR source the principle theoretical limit is given by
half of the repetition rate $R_{\rm rep}$, the factor $1 \over 2$
being inherent to the protocol and realistic values of $R_{\rm
rep}$ reaching up to $100$~GHz. The implementation of the scheme
with bright EPR-entangled beams \cite{silberhorn01} is
experimentally simple and robust and well suited for both
fiber-integrated or free-space transmission.

This work was supported by the Deutsche Forschungsgemeinschaft and
by the EU grant under QIPC, project IST-1999-13071 (QUICOV). The
authors gratefully acknowledge fruitful  discussions with
T.~C.~Ralph, N.~L\"utkenhaus, Ph. Grangier, R.~Loudon, and
S.~Lorenz.


\begin{thebibliography}{10}
\bibitem{gisin01} N.~Gisin, G.~Ribordy, W.~Tittel, and H.~Zbinden,
quant-ph/0101098 (2001).
\bibitem{reid00} M.~D.~Reid, Phys. Rev.  {\bf A 62}, 062308
(2000).
\bibitem{hillery00} M.~Hillery, Phys. Rev.  {\bf A 61}, 022309
(2000).
\bibitem{ralph00a} T.~C.~Ralph, Phys. Rev.  {\bf A 61}, 010303(R)
(2000).
\bibitem{cerf00} N.~J.~Cerf, M.~Levy, and G.~Van Assche, Phys. Rev.  {\bf A 63},
052311 (2001).
\bibitem{ralph00b} T.~C.~Ralph, Phys. Rev.  {\bf A 62}, 062306
(2000).
\bibitem{gottesman00} D.~Gottesman {\it et al}, Phys. Rev.  {\bf A 63}, 022309
(2001).
\bibitem{silberhorn00} Ch.~Silberhorn,  N.~Korolkova, and G.~Leuchs,
"Quantum cryptography with bright entangled beams,"
 IQEC'2000 Conference Digest, QMB6, p. 8, Nice, France, 2000.
 \bibitem{silberhorn}
Ch.~Silberhorn, T.~C.~Ralph, N.~L\"utkenhaus, and G.~Leuchs,
manuscript in preparation.
\bibitem{simon} R. Simon, Phys. Rev. Lett.  {\bf 84}, 2726 (2000).
\bibitem{duan} L.~-M.~Duan, G.~Giedke, J.~I.~Cirac, and P.~Zoller,
Phys. Rev. Lett.  {\bf 84}, 2722 (2000).
\bibitem{reid} M.~D.~Reid and P.~D.~Drummond,  Phys. Rev. Lett.  {\bf 60}, 2731 (1988);
M. D. Reid,  Phys. Rev. {\bf A 40}, 913 (1989). J. Mod. Opt. {\bf
46}, 1927 (1999).
\bibitem{leuchs01} G. Leuchs, Ch. Silberhorn, F. K\"onig, A. Sizmann, and N.
Korolkova, 
in {\it Quantum Information Theory with Continuous Variables},
S.L. Braunstein and A.K. Pati (eds.), Kluwer Academic Publishers,
Dodrecht 2001, in press.
\bibitem{silberhorn01}
Ch.~Silberhorn, P.~K.~Lam, O.~Wei\ss , F.~K\"onig, N.~Korolkova,
and G.~Leuchs, Phys. Rev. Lett.  {\bf 86}, 4267 (2001).
\bibitem{luetkenhaus} N. L\"utkenhaus, Phys. Rev.  {\bf A 54}, 97
(1996); D. {Bru\ss}   and N. L\"utkenhaus, 
quant-ph/9901061 (1999).
\bibitem{ralph01} T.~C.~Ralph, in {\it Quantum Information Theory with Continuous Variables},
S.L. Braunstein and A.K. Pati (eds.), Kluwer Academic Publishers,
Dodrecht, in press, quant-ph/0109096 (2001).
\bibitem{criteria} F. Grosshans and  P. Grangier, Phys. Rev. A {\bf 64},
010301(R) (2001); S.~L.~Braunstein, C.~A.~Fuchs, H.~J.~Kimble,
P.~van~Loock, quant-ph/0012001 (2000).
\bibitem{ou} Z.~Y.~Ou,  S.~F.~Pereira,
H.~J.~Kimble, and K.~C.~Peng,  Phys. Rev. Lett.  {\bf 68}, 3663
(1992).

\end{thebibliography}
\end{document}